# An ultrasensitive device with embedded phononic crystals for the detection and localisation of nonlinear guided waves

**Paweł Kudela[1,*], Maciej Radzienski[1], Marco Miniaci[2], Piotr Fiborek[1], Wieslaw Ostachowicz[1]**

[1] Institute of Fluid-Flow Machinery, Polish Academy of Sciences, Gdansk, Poland

[2] Université de Lille, CNRS, Centrale Lille, Junia, Université Polytechnique Hauts-de-France, UMR 8520 - IEMN - Institut d'Electronique de Microélectronique et de Nanotechnologie, F-59000 Lille, France

## Abstract

In this work, a novel approach for the detection and localisation of nonlinear guided waves often associated with the presence of damage in structural components is proposed. The method is active and consists of a piezoelectric transducer bonded to the inspected structure exciting a narrow frequency band wave packet and sensors placed at the proposed ultrasonic devices with embedded phononic crystals.

[*] Corresponding author.
E-mail address: pk@imp.gda.pl
Phone number: +48 58 6995 251
Address: Fiszera 14 st. Gdańsk 80-231, Poland

Unit cells of phononic crystals are optimized to open a band gap at the excitation frequency so that the excited waves are attenuated, while the sensitivity detection of higher harmonics is increased.

The proposed approach is tested numerically and validated experimentally by considering various manufacturing methods, materials, and unit cell geometries. A parametric study of the angle of attachment of the ultrasonic devices with the embedded phononic crystals to the inspected structure is performed. Band gaps and filtering capabilities of several prototypes are tested.

Numerical simulations of guided wave propagation that include the effect of delamination clapping proved that the proposed designs are sensitive enough to detect higher harmonics by simple signal thresholding. The most promising prototype is tested experimentally showing its capability of detection and localisation of a simulated damage.

## 1. Introduction

Guided waves attracted the interest of many researchers from a mathematical perspective [1], for the associated experimental discoveries [2], for their potential applications in non-destructive testing (NDT) [3] and structural health monitoring (SHM) by using piezoelectric wafer active sensors [4], and for possibilities of their manipulation (filtering [5], focusing [6,7], tunnelling [8] and processing [9]), etc.

In the field of SHM, guided waves are an attractive tool because damage induces anomalies in their propagation such as reflections, mode conversion, velocity, and amplitude change. These anomalies are considered during the development of signal processing methods and algorithms for damage detection, localisation, and size estimation. These methods usually assume the linear wave response at sensing points that are used for the development of damage indexes. Many promising damage identification methods utilizing linear guided waves have been developed [10–13]. However, recent trends focus on pinpointing damage at an even earlier stage of growth by using nonlinear guided waves.

Non-linearity in guided wave propagation is induced mostly by two factors: (i) non-linear material behaviour and (ii) contact interactions. Non-linear material behaviour can be described by constitutive relations in which strain-displacement and stress-strain response have non-linear character [14]. The other type of nonlinearity comes from the abrupt changes in stiffness at material interfaces coming into and out of contact [15]. Recent findings indicate that contact acoustic nonlinearity (CAN) constitutes a non-classical wave nonlinearity that arises from the interaction between acoustic waves and materials at interfaces that produces noticeable changes in amplitude and frequency response of registered signals recognised as sub- and super-harmonics or higher-order wave modes [16,17]. The study of CAN plays a key role in the early detection of cracks, debonding, or delaminations due to the variance of propagating guided waves at damage-affected locations. Therefore, several numerical modelling methods were employed to study damage-induced nonlinear effects. Van Den Abeele et al. [18] used a multiscale approach in which a non-linear hysteretic stress-strain relation was implemented into a numerical elastodynamic finite integration. Joglekar and Mitra [19] studied a breathing crack in a 1D waveguide by using bilinear stiffness characteristics embedded into Fourier spectral finite elements. They investigated nonlinear frequency mixing resulting from the modulation effects induced by the breathing crack. Pecorari and Solodov [16] have shown that an interface between asperities in Hertzian contact is characterized by a response in which the second harmonic component dominates the spectrum of the nonlinear scattered waves. Moreover, mechanisms such as partial slip and adhesion produce a cascade of higher-order harmonics, with the odd harmonics dominating over the even ones. Nonlinear modulation of Lamb modes by clapping delamination in bi-layer structure was studied in [15]. A simplified approximate quasi-stationary approach combined with modal decomposition was used to determine the delamination closing criterion based on the normal displacement induced by the slow force or wave. Much more advanced modelling techniques based on the time domain spectral element method, also called spectral finite element method have been explored in recent years [20–22].

Soleimanpour et al. [17] used not only numerical but also experimental approaches for detecting and locating cracks in isotropic plates using the second harmonic of the S0 Lamb wave mode induced by cracks. Higher harmonics were also used for detecting other damage types such as disbond in aluminium stiffened panel [23], delamination in woven fibre-reinforced composite laminate [24] and even corrosion damage on the aluminium panel [25] proving that this is a viable option for SHM.

Nevertheless, several issues arise in practical implementation due to the very small amplitude of higher harmonics, which are often buried in noise (the amplitude of higher harmonics is generally one or two orders of magnitude lower than excitation frequency). Therefore, powerful amplifiers are required to drive piezoelectric transducers. Such voltage amplifiers often operate beyond their linear range. It becomes challenging to distinguish between inherent nonlinear effects introduced by the measurement setup (equipment and actuators in active SHM) and structural discontinuities (e.g. fatigue cracks and delaminations).

One of the solutions to the problem of nonlinear effects induced by permanently attached to the monitored structure actuators is the application of a fully non-contact measurement setup, such as based on lasers [26]. However, such a solution is difficult to implement in the field. Another option, investigated in [27,28], involves surrounding the actuator with a phononic crystal (PC) waveguide suppressing higher harmonics so that only guided waves of desired frequencies are excited.

Phononic crystals are periodic structures exhibiting unconventional dynamic properties enabling the manipulation of guided and acoustic waves. The periodic variation of the material properties of PCs can be designed in such a way that band gaps are formed, i.e. frequency range for which the propagation of the elastic waves through the PC is strongly attenuated. The band gap phenomenon is caused by constructive/destructive reflection and superposition of waves at the interface of periodic heterogeneities, i.e., Bragg resonant scatterings. It gives unique opportunities in the design of analogue filters with potential applications in radio frequency communications and acoustic imaging for medical ultrasound and NDT testing [29]. At the other end of the frequency spectrum, the application of seismic barriers emerges [30]. However, the potential of metamaterials and PCs in SHM is still in its infancy, and only a limited number of concepts of PC waveguide transducers for nonlinear elastic wave sensing, filtering and localisation of the source of nonlinear waves have been proposed [27,28,31], so far.

Based on previous studies, we explore the potential application of PCs for SHM. We target the problem of small amplitudes of guided waves related to damage-induced higher harmonics. The focus is on the sensing side so that the proposed filtering device can complement the solutions proposed in [27,28]. This problem is tackled both from a numerical modelling perspective as well as manufacturing and experimental validation. The aim is to develop an ultrasensitive device which can be mounted anywhere on the structure. Moreover, due to analogue filtering capabilities, it is possible to use it directly as an event-triggered sensor without any digital filtering [32]. Another option is to combine it with tiny machine learning (TinyML) for inferencing on edge devices [33].

## 2. Spider inspiration

The proposed ultrasonic device is inspired by spiders, which are among the most vibration-sensitive living creatures. They possess extremely efficient strain detectors known as lyriform organs, capable of transducing mechanical loads into nervous signals embedded in their exoskeleton [34]. It is a slit organ in which each slit is of different dilatation, length, and angle.

It allows for the sensing of vibration at a wide range of frequencies and wavelengths. Spiders can process these signals and sense a prey trapped on the spider's web. The sensing capabilities of spiders have driven the design of bio-inspired solutions in terms of sensor technology reviewed in [35] including an ultrasensitive mechanical crack-based sensor [36].

Our aim is not biomimicry, i.e. a pure replication of the mechanism behind lyriform organs, but rather getting inspired in the design of our ultrasonic devices by the fact that spiders' legs have embedded filtering and sensing capabilities. The filtering mechanism is going to be realized through PCs, whereas sensing through widely available piezoelectric transducers. Contrary to sensing vibrations of frequencies below 1 kHz as in the case of spiders' legs, we are going to rely on ultrasonic guided waves of frequencies in the range of 50-500 kHz to detect and localise defects in the structure, analogously to spiders sensing preys on their webs. The proposed spider-leg-inspired design allows for omnidirectional sensing and represents a reasonably lightweight structure, potentially portable at the same time.

*2.1 General concept*

Defects cause anomalies of propagating guided waves, such as wave reflections and mode conversions (linear effects) and a generation of higher harmonics (non-linear effects). These anomalies can be potentially used for damage identifications. The general concept of damage detection by utilizing higher harmonics is depicted in Fig. *1*. The wave packet of carrier frequency $f_1$ is excited at point **A**. The propagating wave encounters a defect, which becomes the source of higher harmonics generated, e.g. due to delamination clapping or breathing of fatigue crack. The guided wave propagates further to point **B,** where frequency components are around primary frequency $f_1$, and higher harmonics are represented by $f_2$ and $f_3$. Next, the propagating wave reaches the ultrasonic device that has embedded PCs responsible for filtering out the main frequency harmonic $f_1$. In consequence, only higher harmonics induced by defect are registered at point **C**. By setting a threshold, it is possible to distinguish between the healthy state and the damaged state of the structure.

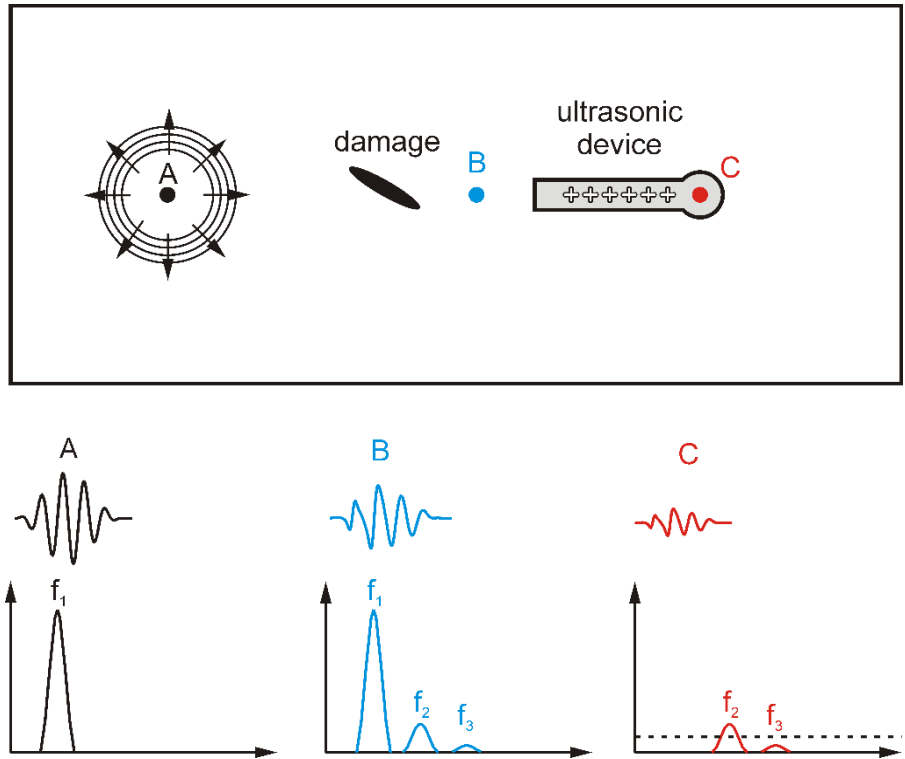

Fig. 1. Metamaterial-aided defect detection by utilization of higher harmonics.

## 3. Design considerations and technological challenges

The design of the ultrasonic device requires a lot of parameters to be considered including material selection, geometry, manufacturing technology, frequency range, etc. Since it was an iterative process, several designs were tested and are described here, including failed ones, to underline the lessons learned along the way. The parameters which were kept the same are the unit cell of PC side length of 8 mm and thickness of 3 mm. The following materials were considered: aluminium alloy, polymer resin and steel alloy. The respective material properties used in numerical simulations are given in Table 1.

Table 1. Material properties.

| Material | Young's modulus [GPa] | Poisson's ratio | Density [kg/m$^3$] |
|---|---|---|---|
| Aluminium alloy | 71 | 0.33 | 2700 |
| VeroTM polymer | 2.96 | 0.38 | 1180 |
| AISI 304 steel | 190 | 0.27 | 7955 |

*3.1 Consideration of manufacturing methods*

The following manufacturing methods can be considered for producing the ultrasonic device with embedded PCs:

a) Computer Numerical Control (CNC) machining,

b) Electrical Discharge Machining (EDM),

c) 3D printing,

d) Laser cutting.

Due to the relatively small dimensions and complex shapes of the cavities in the analysed PCs, a micro-CNC machine with tiny milling cutters is required rather than widely available CNC machines. However, by using micro-CNC, it is still difficult to achieve a precision of 0.1 mm, especially for materials thicker than 2 mm. In such a case, milling cutters are exposed to high bending moments and tend to break. The EDM drilling method is superior in this regard as it can drill holes with diameters as small as 0.065 mm and a depth of up to 1 m. However, wire EDM must be used for machining complex shapes which makes it cumbersome when multiple cavities are needed because a pilot hole must be created for each hole and wire threaded through it. EDM is limited to drilling/spark machining conductive materials. It is a relatively expensive process because of the time it takes to produce parts and the high energy consumption.

We have tried to use 3D printing and laser cutting for manufacturing a prototype of the ultrasonic device as a compromise of accuracy, cost, manufacturing time, and availability of these technologies at our research institutes. Additive manufacturing technologies vary significantly depending on whether plastics, liquids, or powder grains are used. A failed example of one of our prototypes by using an EOSINT M 280 3D printer and Direct Metal Laser Sintering (DMLS) of aluminium AlSi10Mg powder is shown in Fig. 2. The main issue is that the support material is required under the whole spider, which next is mechanically cut off from the base. Unfortunately, the cutting process caused some legs to break. It can be seen in Fig. 2 that the support material, of a slightly brighter colour, still needs to be removed.

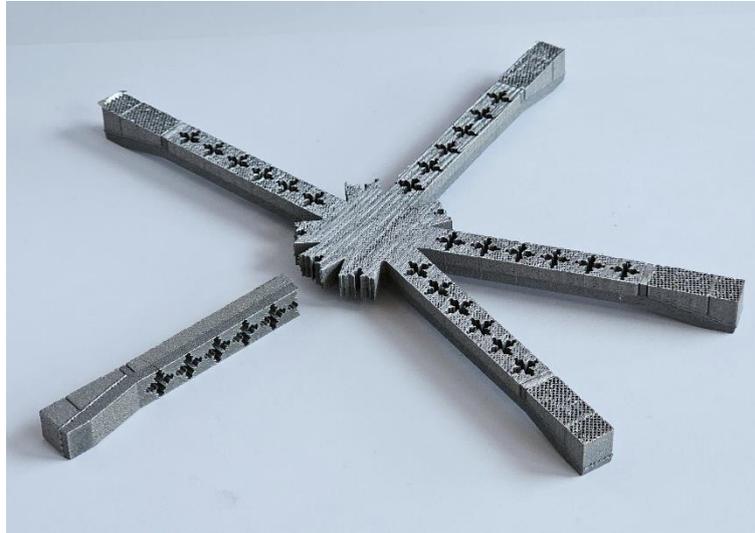

Fig. 2. A spider-inspired ultrasonic device manufactured by using the DMLS method.

Contrarily, 3D printing using polymer resin and Masked Stereolithography (mSLA) or PolyJet printing technology enables precise enough printing and allows easy removal of the printing part from the base.

Another manufacturing method which can be used is laser cutting. The tolerance range of the laser is from 0.003 mm to 0.006 mm. It is also a relatively low-cost manufacturing method. A significant downside to laser cutting is the evaporation of material, especially plastics, which produces harmful fumes and gases. Another limitation is that the laser path is defined in 2D, hence it is usually applied to materials in the form of sheets. The spider-inspired geometry can be laser-cut from a metal sheet, but spiders' feet have to be bent and ground to ensure even contact with the surface of an inspected structure.

Three prototypes that have been investigated are presented in Fig. 3. The first prototype was a spider-inspired device laser-cut from an aluminium alloy sheet (Fig. 3a) which has 8 legs with 4-unit cells in each leg. It can be noticed that the legs are very fragile due to the small amount of material surrounding cavities embedded in the legs and material evaporation during laser cutting. To rectify this problem, the topology optimization described in [37] was utilised to obtain a cavity shape with enough structural stiffness suitable for manufacturing while maintaining a relatively large band gap. This resulted in a 3D-printed prototype shown in Fig. 3b. A Stratasys Objet350 Connex3 printer along with PolyJet printing technology was used for printing it. A spider-like geometry made of VeroTM thermoplastic polymer (see Tab. 1) has 9 legs for better omnidirectional sensitivity. The third iteration of the prototype, shown in Fig. 3c, was simplified to laser-cut steel legs. Two versions were made with one and two rows of unit cells. Two rows of unit cells lead to a stiffer structure but a smaller band gap. Since the version with one row of unit cells is stiff enough and results in a wider band gap, it was used for further investigations. It can be noted that the end of the leg was machined at an angle of 12° so that the leg could be bonded to the inspected structure. The reasoning behind selecting this angle is explained in the next section.

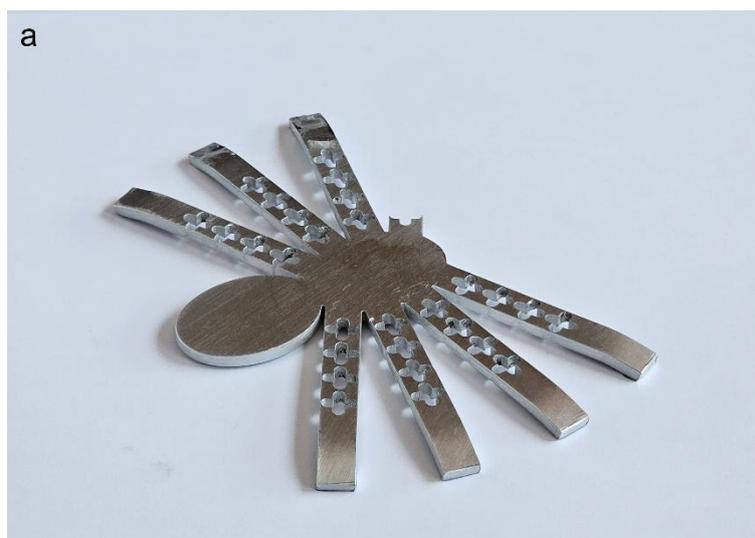

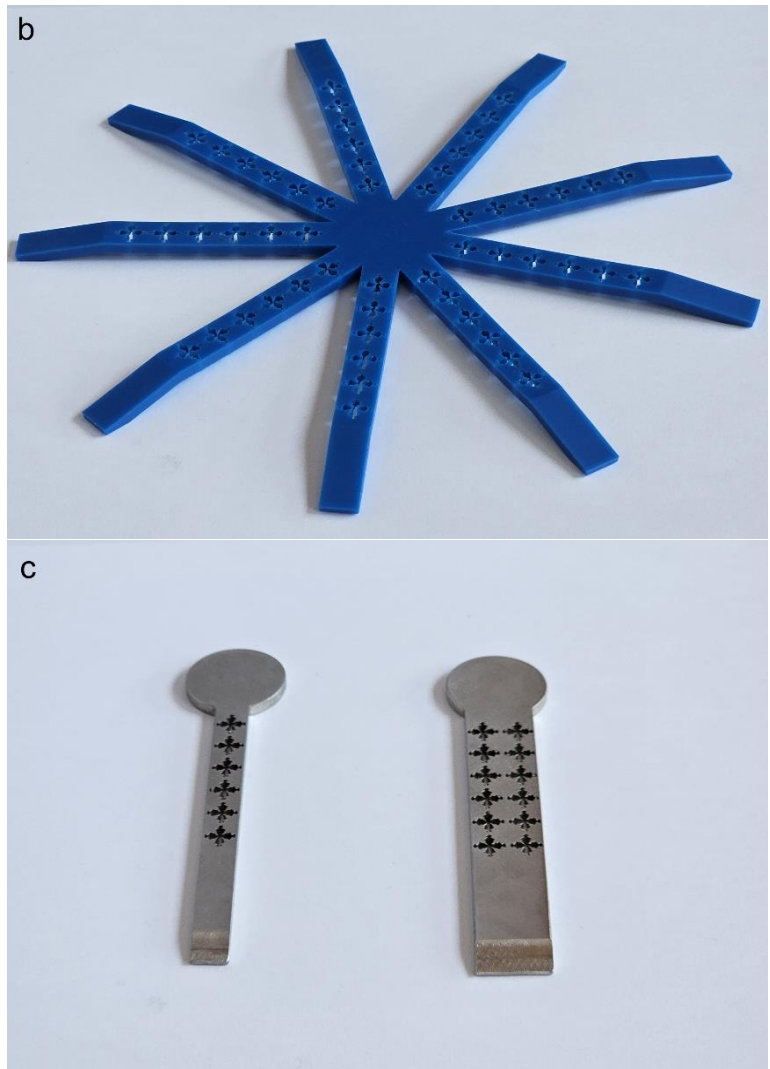

Fig. 3. Prototypes of ultrasonic devices: a) spider-inspired device laser-cut from aluminium alloy sheet, b) 3D-printed polymer spider-like geometry, c) laser-cut steel legs.

*3.2 Parametric studies of foot angle*

The prototyping process was supported by parametric studies conducted numerically using COMSOL Multiphysics software. The angle $β$ of the foot with respect to the surface of the plate, as indicated in Fig. 4, was investigated considering the cross-section through the waveguide without PCs. The excitation was applied at point A and sensing was conducted at points $S_1$ (on the leg) and $S_2$ (on the surface of the plate). The energy $E_S$ of registered signals was calculated for angles ranging from 0° to 90° in 1° increments. The results are shown in Fig. 5. The signal energy at both locations is on a similar level for angles in the range of 3-20°. The aim was to obtain as high signal energy as possible at point $S_1$ (on the leg) without an excessive foot length. As a compromise, a 12°-foot angle was selected for further studies and prototype designs.

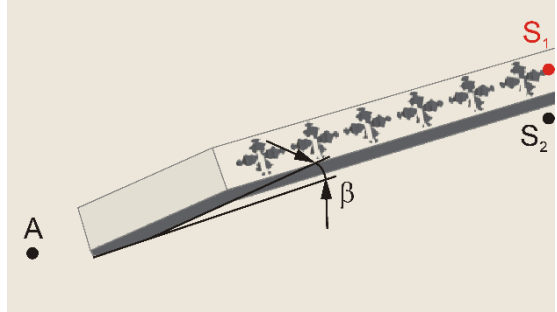

Fig. 4. Parametric studies of guided wave actuation at point A and sensing at points $S_1$ and $S_2$.

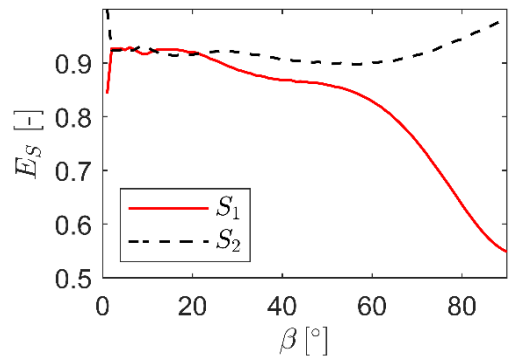

Fig. 5. Signal energy $E_S$ dependence on the angle $\beta$ for signals transmitted to point $S_1$ on the leg and $S_2$ on the surface of the plate, respectively.

*3.3 Parametric studies of the number of unit cells*

Numerical tests with isolated legs were performed in which the number of unit cells varied from 3 to 6. The excitation of Hann-windowed sinusoidal signal with a carrier frequency of 56 kHz and 5 cycles was applied at one side of the waveguide, whereas the sensing point was at the other side of the waveguide. Transmission coefficients as a function of the number of unit cells in the waveguide were calculated using the following formula:

$$T = 10 \log_{10}\left(\frac{E_{out}}{E_{in}}\right),$$

where $E_{in}$ – initial energy of excited waves, $E_{out}$ – the amount of energy transmitted across the PC region.

The results are gathered in Table 2. A satisfactory drop in the signal energy transmission is observed for the case of 4-unit cells, however, for SHM application in which higher harmonics of tiny amplitudes must be detected, 6-unit cells may be required.

Table 2. Signal energy drops depending on the number of unit cells.

| Unit cells | 3 | 4 | 5 | 6 |
|---|---|---|---|---|
| Transmissibility | -20.6 dB | -25.7 dB | -29.8 dB | -32.1 dB |

*3.4 Band gaps*

Two geometries of unit cells with a side length of 8 mm and thickness of 3 mm were considered as shown in Fig. 6. The geometry with a cross-like shape cavity presented in Fig. 6a is the same as proposed in [31]. This geometry was used in a laser-cut aluminium spider (see Fig. 3a). The geometry presented in Fig. 6b was obtained through topology optimization as proposed in [37], where both structural rigidity and band gap size were accounted for. It should be noted that the minimal wall thickness in both unit cells is similar (400 µm and 391 µm, respectively), but there is overall more material in the latter unit cell, resulting in greater stiffness when unit cells are periodically arranged in a 1D structure. This geometry was used in a 3D-printed polymer spider (see Fig. 3b) and laser-cut steel legs (see Fig. 3c). Therefore, a total of three cases were investigated. For each case, numerical calculations and experimental measurements were conducted to confirm band gaps.

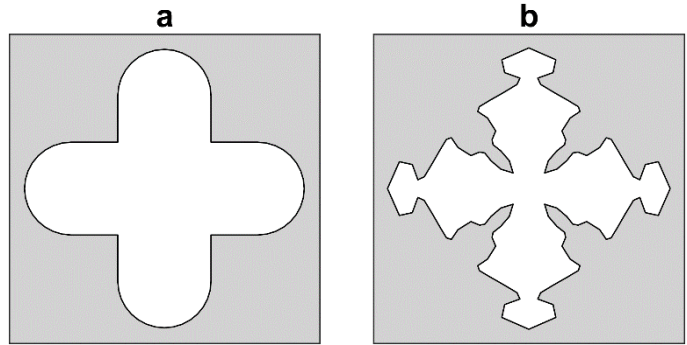

Fig. 6. Unit cell geometries with grey representing solid material and white representing the cavity: a) cross-like shape used in laser cut aluminium spider (see Fig. 3a), b) optimised cavity shape used in 3D-printed polymer spider (see Fig. 3b) and laser cut steel legs (see Fig. 3c).

COMSOL Multiphysics software was used for calculating the dispersion diagrams of 1D periodic structures composed of unit cells presented in Fig. *6* and appropriate material properties listed in Tab. 1. Unit cells were discretized into solid triangular prism mesh elements and the eigenfrequency solution was obtained through the finite element method. Bloch periodic boundary conditions were applied to the unit cell along one direction whereas perpendicular faces remained free. The dispersion relation $\omega(k)$ was solved by varying the wave vector $k$ along the wave propagation direction. Since it is the 1D arrangement of cuboid unit cells, the focus is on the wave propagation along Γ-X direction. The numerical results are presented on the left side in Fig. *7*-Fig. *9*. The *x*-axis represents the reduced wave vector $k^*$ and the *y*-axis represents the frequency *f*. Moreover, colour-coded polarizations were applied to distinguish between in-plane and out-of-plane modes [5]. Namely, colours close to dark red indicate vibration modes that are predominantly polarized out-of-plane, while colours close to blue indicate modes that are predominantly polarized in-plane. The main band gap of interest and secondary band gap at higher frequencies were highlighted in grey colour.

Because the number of unit cells in the analysed three prototypes (see Fig. *3*) is finite and the applied materials have significantly different damping properties, experimental validation was necessary to confirm band gaps and conclude about the potential usefulness of the designed prototypes. The experiment was performed using a piezoelectric actuator in the form of a disk with a diameter of 10 mm and thickness of 0.5 mm and a scanning laser Doppler vibrometer (Polytec PSV-400). The piezoelectric actuator was bonded to the central part of each prototype and was excited by the broadband chirp signal in the frequency range of 0-512 kHz and lasting 12.8 ms. The response was acquired by the laser vibrometer at a grid of points on the leg and then averaged. The energy of the guided wave propagation signal transmitted through PCs is shown on the right side in Fig. *7*-Fig. *9*.

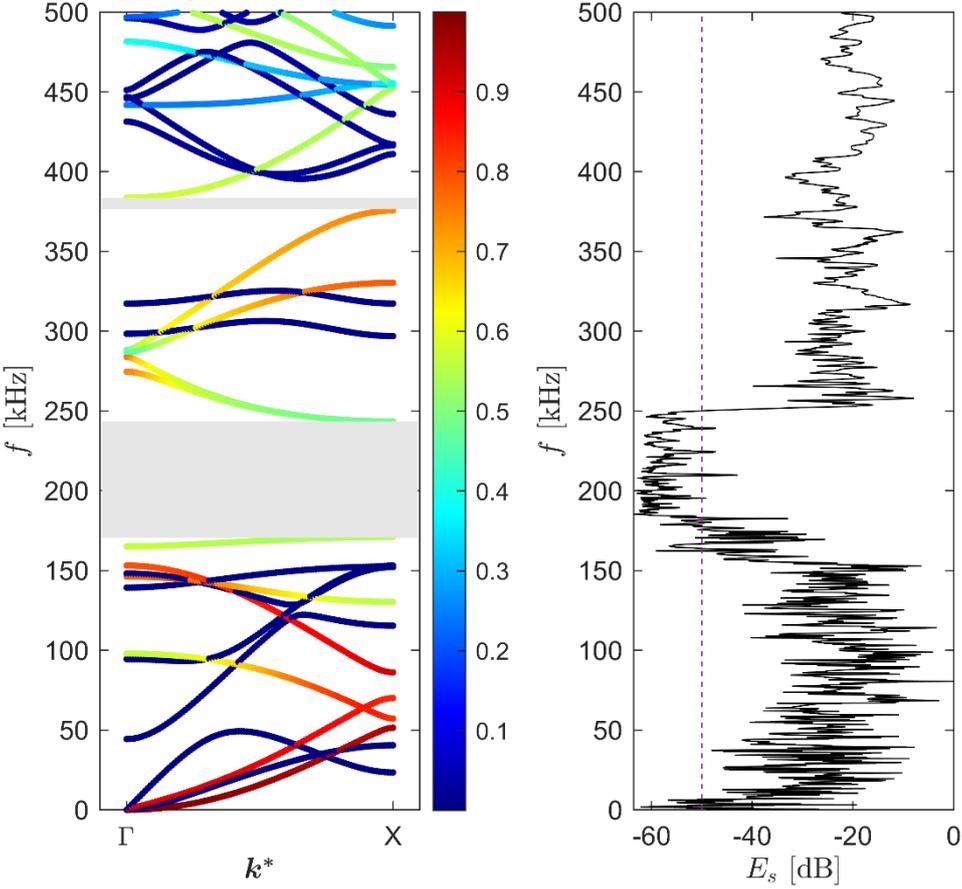

Fig. 7. The band gap in the laser-cut aluminium spider's leg: numerical and experimental analysis.

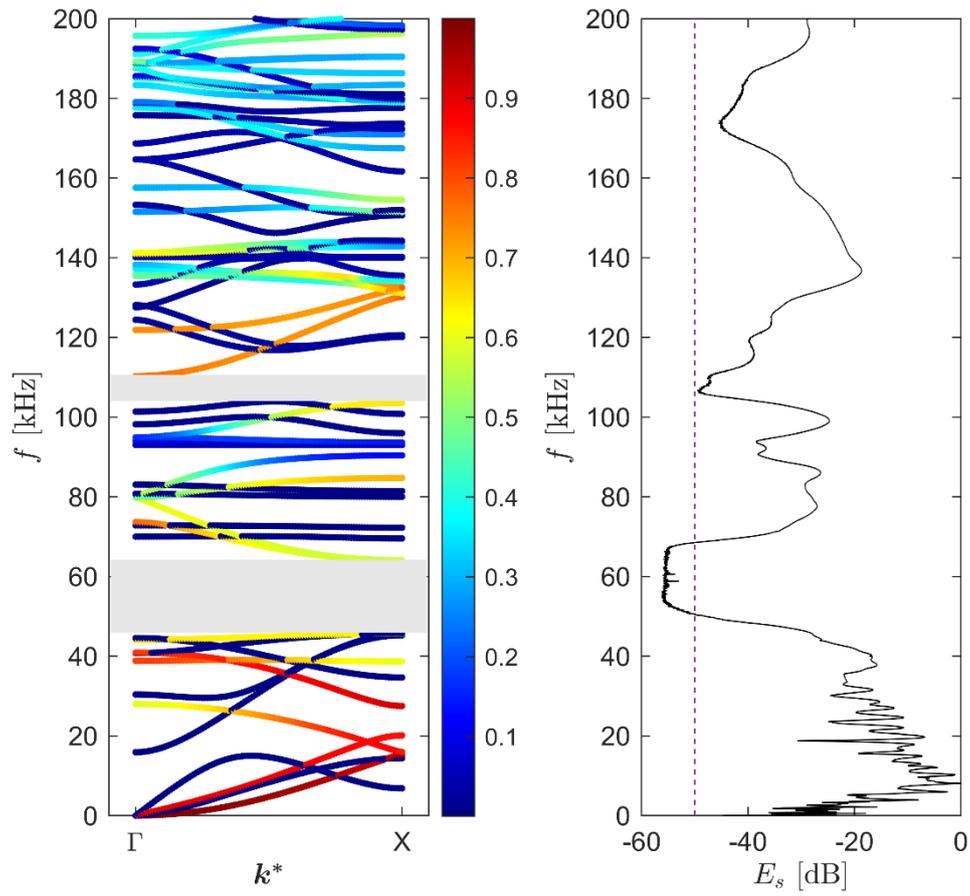

Fig. 8. The band gap in the 3D-printed polymer spider leg: numerical and experimental analysis.

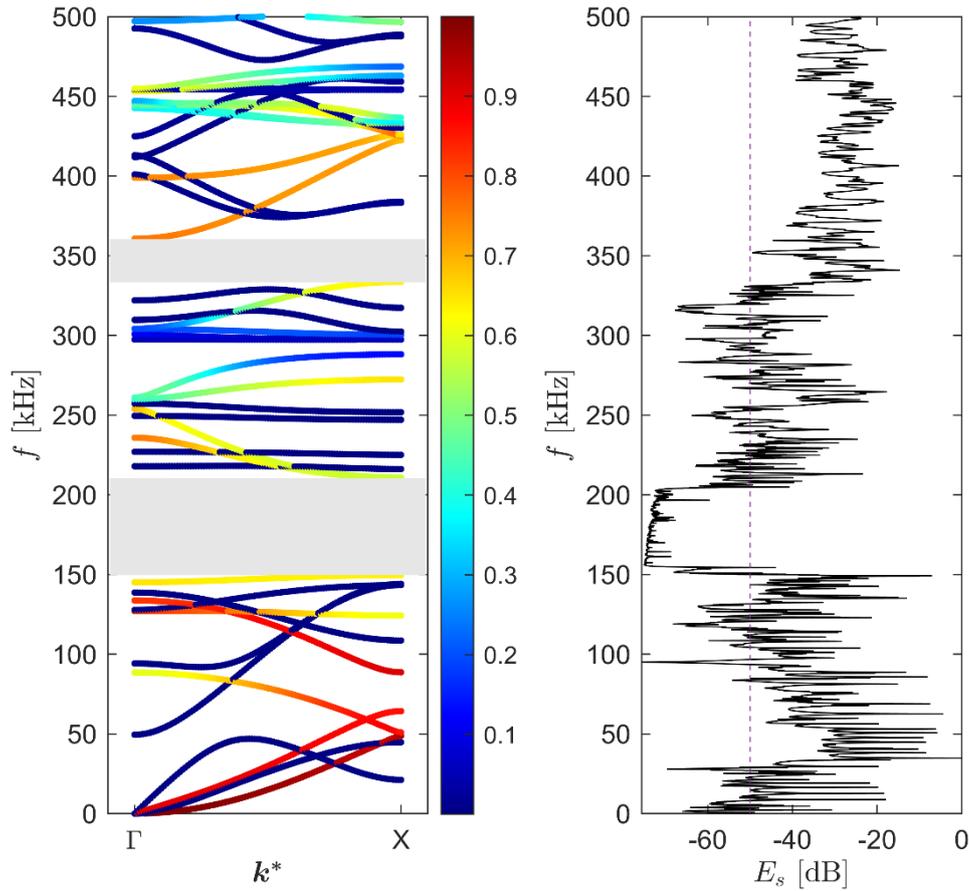

Fig. 9. The band gap in the laser-cut steel spider's leg: numerical and experimental analysis.

A very good match between the numerical calculations and experimental measurements of the band gap frequency range was obtained in all three cases. The band gap characteristics are summarized in Tab. 3. The largest relative gap to mid-gap ratio, denoted by BG%, equal to 30.9% was achieved for the laser-cut aluminium spider's leg. However, the largest signal energy drop in the band gap is observed in the laser-cut steel spider's leg. It is much lower than the indicative dashed line placed at -50 dB, for which the band gap was estimated. It is also evident that signal energy at higher frequencies is considerably attenuated for the 3D-printed polymer spider leg in comparison to metallic prototypes. For the general concept described in section 2.1, it is crucial that if the excitation frequency is at the mid-gap frequency $f_m$, the signal energy level difference $E_s(2f_m)$-$E_s(f_m)$ or $E_s(3f_m)$-$E_s(f_m)$ should be as high as possible.

Experimental measurements indicate that the best prototype in that regard is the laser-cut steel spider's leg with a signal energy level difference equal to 49.1 dB.

Table 3. Band gap parameters: $f_m$ - mid gap frequency, BG% - relative gap to mid-gap ratio and signal energy level difference $E_s(2f_m)$-$E_s(f_m)$, $E_s(3f_m)$-$E_s(f_m)$.

| Prototype | Numerical sim. | | Experimental measurements | | | |
|---|---|---|---|---|---|---|
| | $f_m$ [kHz] | BG% | $f_{m\text{-}50dB}$ [kHz] | BG%$_{\text{-}50dB}$ | $E_s(2f_m)$-$E_s(f_m)$ [dB] | $E_s(3f_m)$-$E_s(f_m)$ [dB] |
| Fig. 3a | 205.8 | 33.7 | 216.6 | 30.9 | 35.3 | - |
| Fig. 3b | 54.6 | 34.4 | 59.5 | 30.3 | 16.2 | 13.6 |
| Fig. 3c | 180.3 | 33.9 | 179.6 | 27.9 | 49.1 | - |

## 4. Proof of the concept

Proof of the concept of an ultrasensitive ultrasonic device capable of detecting nonlinear waves emanating from damage was realized numerically and experimentally.

*4.1 Numerical verification*

The proposed model is based on the parallel implementation of the time domain spectral element method [38], expanded by electromechanical coupling to cover piezoelectric transducers. The delamination clapping mechanism in the in-house code [39] was implemented similarly to the works of Yu et al. [22] and Li et al. [40]. The structure, in the form of a plate, was divided into two layers joined by a matching interface except for the damaged area. The penalty method was adopted to calculate forces between surfaces in contact. The frictionless model was used, assuming that the tangential forces are negligible with respect to the normal forces. The contact forces depend on the amount of node penetration according to the formula:

$$F_{ci} = \epsilon_N H(-g_{Ni})g_{Ni}A_i \tag{1}$$

where $\epsilon_N$ is the penalty parameter, $A_i$ is the vector associated with the contact point $i$, $H(-g_{Ni})$ is Heaviside step function, and $g_{Ni}$ is the gap between contact nodes $i$. It should be noted that in our approach the contact forces are calculated iteratively by adjusting the penalty parameter. The resulting equation of motion has a form:

$$\mathbf{M}\ddot{\mathbf{u}} + \mathbf{C}\dot{\mathbf{u}} + \mathbf{K}\mathbf{u} + \mathbf{F}_c + \mathbf{F}_{int} = \mathbf{F}_e \tag{2}$$

where $\mathbf{M}$, $\mathbf{C}$, $\mathbf{K}$ are the mass, damping and stiffness matrices, respectively, $\mathbf{F}_e$ is the vector of external forces, $\mathbf{F}_{int}$ is the vector of interface forces, necessary to join structural components. It should be noted that the mesh can be prepared so that the nodes of combined components coincide. In such a case a matching interface can be used. For the time integration of Eq. (2), the central difference method was used.

The simulations were performed for two bonded sheets of aluminium alloy plates with thicknesses of 2 mm and 1 mm, respectively, and planar dimensions of 280 mm by 280 mm. It was assumed that they are perfectly bonded except for the delamination area of elliptic shape (semi-minor axis length 5 mm, semi-major axis length 10 mm). The delamination centre was located 56 mm from the edges of the lower left corner of the plate.

Two cases were simulated in which the ultrasonic device was in the form of a leg made of polymer and steel, respectively (see the mechanical properties in Tab. 1) with 6 unit cells of geometries shown in Fig. 6b. The signal in the form of Hann-windowed sine signal of carrier frequency 56 kHz for polymer device and 180 kHz for steel device was applied to the piezoelectric transducer located in the centre of the plate. In both cases to ensure that the frequency content of the excited signal fits in the band gap range, a relatively narrow frequency band signal with 35.5 cycles was selected. The piezoelectric actuator and sensor had the form of discs of diameter 5 mm and a thickness of 0.5 mm.

A close-up of the device with the actuator and sensor is shown in Fig. 10a. It should be noted that the render of the ultrasonic device is clipped to show the actuator, which is bonded to the surface of the plate underneath the device. The ultrasonic guided wave propagation at selected time frames for the case of polymer device is shown in Fig. 10. The interaction of guided waves with delamination is visible in Fig. 10d. It is also evident that the ultrasonic device with embedded PCs blocks perfectly excited waves, not allowing them to reach the sensor (see Fig. 10e).

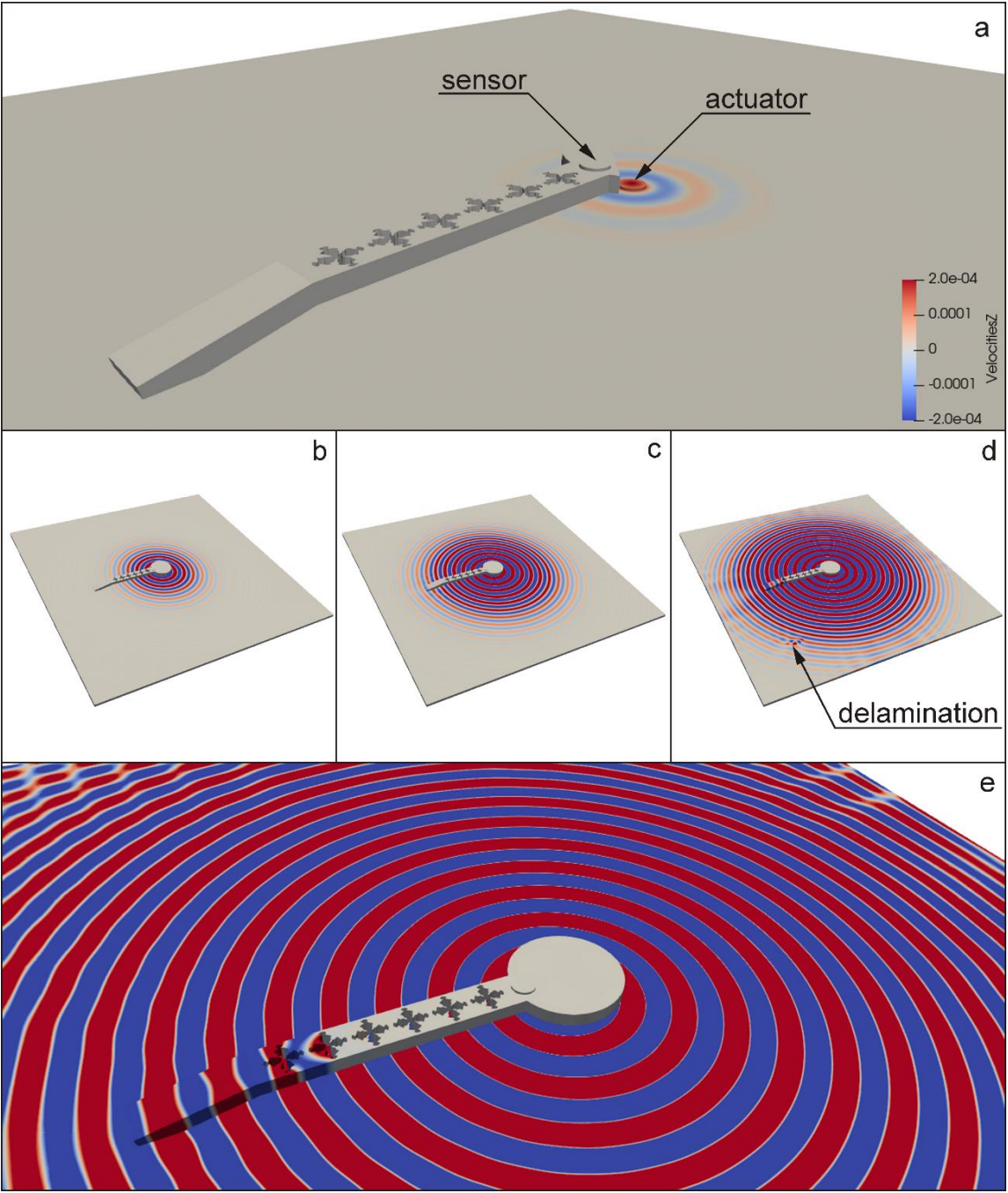

Fig. 10. Ultrasonic guided wave propagation in a plate with a delamination and ultrasonic device with embedded PCs at the time frames: a) 31.2 µs, b) 46.9 µs, c) 140.6 µs, d) 187.5 µs, e) 250 µs.

The detailed analysis of signals registered by the sensor on the ultrasonic device is shown in Fig. 11 and Fig. 12. The reference case (without damage) and the case with delamination are shown. The ultrasonic device made of polymer and designed for lower frequencies is not as effective as that made of steel. The second harmonic at 112 kHz is slightly lower than the excited waves at 56 kHz (see Fig. 11) which does not comply with the purpose of the device, however, the third harmonic at 168 kHz can be used as an indication of CAN due to delamination clapping. On the other hand, the ultrasonic device made of steel designed for higher frequencies is capable of isolating higher harmonics at 360 kHz and 540 kHz (see Fig. 12).

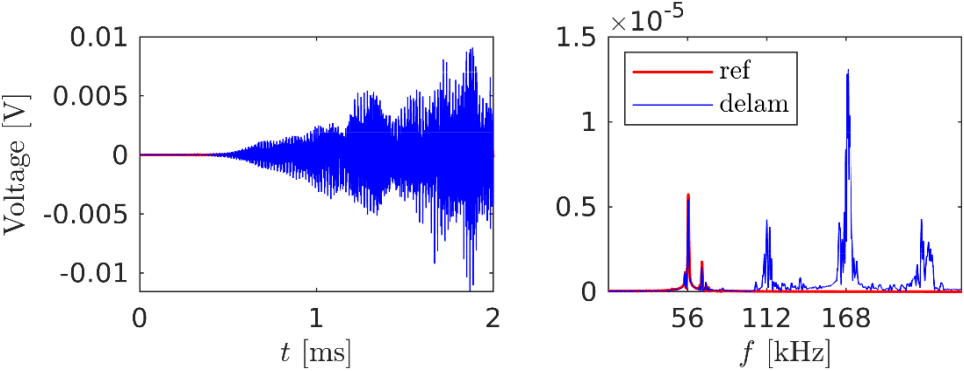

Fig. 11. Time and frequency domain signal response at the sensor on the ultrasonic device made of polymer.

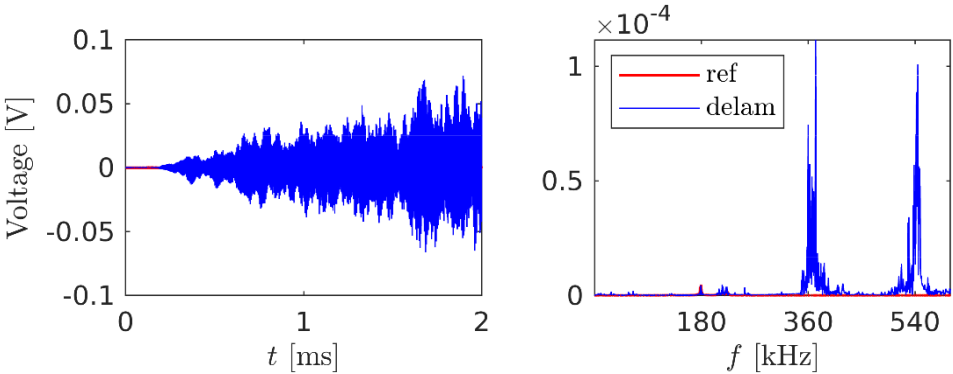

Fig. 12. Time and frequency domain signal response at the sensor on the ultrasonic device made of steel.

*4.2 Experimental validation*

Experimental validation was performed for the case of the ultrasonic device made of steel. Four devices in the form of laser-cut steel legs were bonded using cyanoacrylate glue to the aluminium plate of thickness 2 mm at random positions near the centre of the plate (see Fig. 13). The schematic spatial arrangement of the devices on the surface of the plate is presented in Fig. 14 whereas the detailed coordinates of elements such as centres of feet contacting the plate, actuator and simulated damage positions are given in Tab. 4. The coordinates are given in relation to the origin at the lower left corner of the plate.

The experiment was conducted according to the scheme given in Fig. 14. MATLAB and Instrument Control Toolbox were utilised to control devices. The waveform generator (NI PXIe-5413) generated two waveforms, namely Hann windowed sine signals at carrier frequencies at the first harmonic equal 180 kHz and the second harmonic equal 360 kHz, with 35.5 and 71 cycles, respectively. The first signal was amplified to 20 V by using the Krohn Hite 7500 amplifier and kept at this value for all measurements, while the second signal amplification varied from 0.1 V to 2 V with the step of 0.1 V. The amplification was realized by using the EPA-104 Piezo Linear Amplifier. The objective was to simulate the non-linear effect of higher harmonics at four locations, which have much smaller amplitude than the first harmonic. The higher harmonic signal was appropriately delayed to simulate the wave mixing effect. The mixed guided waves propagated to sensor numbers 1-4 located on ultrasonic devices, then were amplified by a charge amplifier (CEDRAT LWDS system) and acquired by an oscilloscope (NI PXIe-5105). Fast Fourier transform of registered signals reveals the magnitudes of $A_{f1}$ and $A_{f2}$, at the first and the second harmonic, respectively.

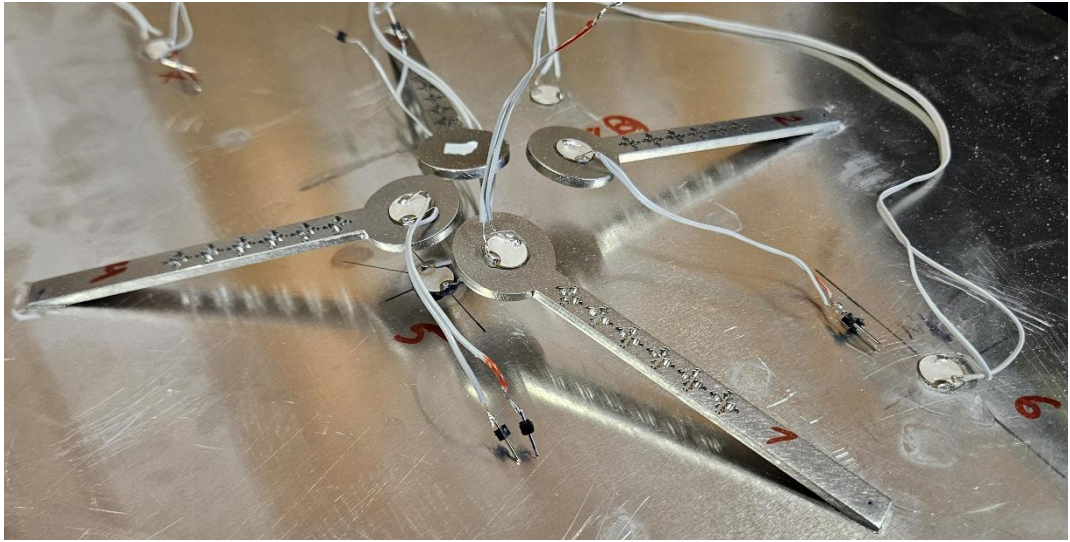

Fig. 13. Four ultrasonic devices bonded to an aluminium plate.

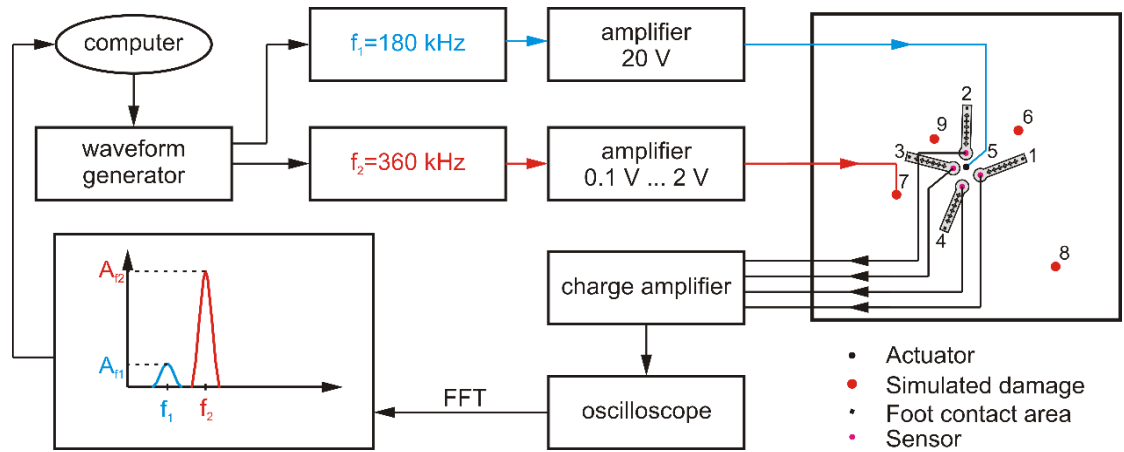

Fig. 14. Scheme of the experimental setup.

Table 4. Coordinates of experimental setup elements with origin at the lower left corner of the plate.

| No | Element name | Coordinates | |
|---|---|---|---|
| | | $x$ [mm] | $y$ [mm] |
| 1 | Foot | 352 | 292 |
| 2 | Foot | 253 | 384 |
| 3 | Foot | 145 | 296 |
| 4 | Foot | 206 | 165 |
| 5 | Actuator | 250 | 250 |
| 6 | Simulated damage (a) | 337 | 334 |
| 7 | Simulated damage (b) | 115 | 226 |
| 8 | Simulated damage (c) | 396 | 105 |
| 9 | Simulated damage (d) | 199 | 320 |

The results of the varied voltage ratio of excited signals applied to the selected simulated damage and signal of constant voltage applied to the actuator, respectively, are presented in Fig. *15*. The magnitude at the first harmonic $A_{f1}$ remained constant as expected. The magnitude of the second harmonic $A_{f2}$ started to become higher than the magnitude of the first harmonic at an excitation voltage ratio of 0.03. This means that thanks to the filtering capabilities of the designed ultrasonic device it is possible to detect higher harmonics at the amplitude level more than 30 times lower than the first harmonics without any signal processing. It should be added that a similar experiment was repeated for the case of the third harmonic but it was unsuccessful since the amplitudes of the third harmonic were very small and remained below the amplitude of the first harmonic for an excitation voltage ratio up to 0.1.

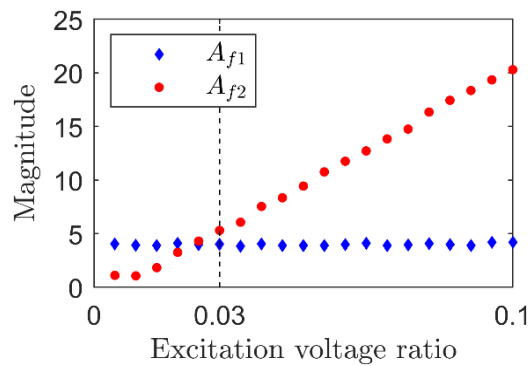

Fig. 15. Effectiveness of the nonlinearity detection by the proposed ultrasonic device.

The registered signals were processed further for qualitative assessment of the simulated damage localisation. The times of arrival were estimated for simulated damage positions as listed in Tab. 4. The signals with the times of arrival indicated by red vertical lines are shown in Fig. 16. For a more precise analysis, band-pass filtering around the second harmonic is necessary to isolate the relevant frequency components. The Hilbert transform can also help automate the process by providing an envelope of the signal, making it easier to identify the arrival times. Additionally, the ultrasonic device's attachment to the plate offers access to both sides of the device. This dual access allows for the possibility of attaching piezoelectric transducers on both sides, enabling the separation of symmetric and antisymmetric modes of guided waves. This separation can facilitate a more accurate estimation of the time of arrival, although this approach was not tested in the current study.

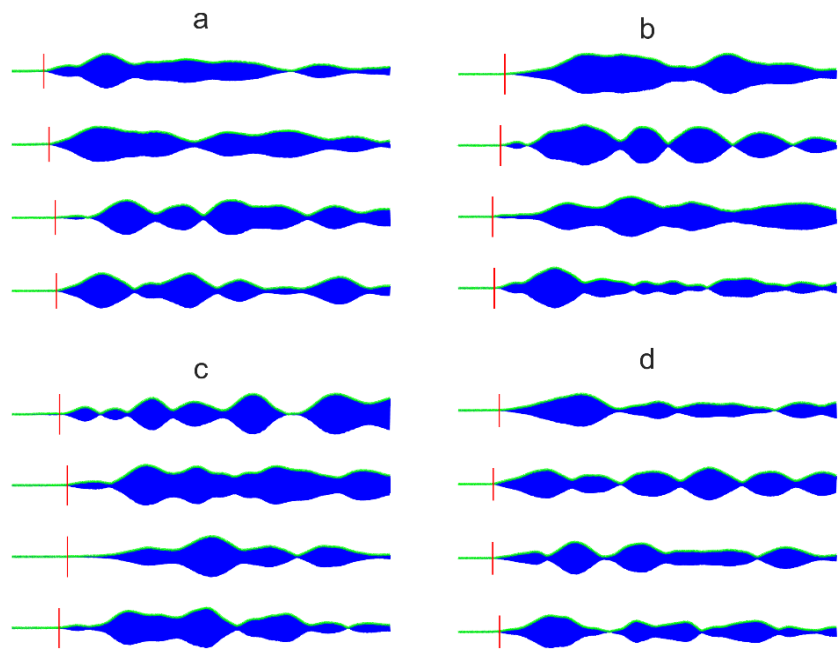

Fig. 16. Estimated time of arrivals (red lines) in signals with applied Hilbert envelope (green) registered at four legs (blue): a) simulated damage at position 6, b) simulated damage at position 7, c) simulated damage at position 8, d) simulated damage at position 9.

Once times of arrival are known, two types of damage localisation algorithms can be applied: based on the difference of the arrival time (hyperbole method) and based on the time of flight from the actuator to damage and from the damage to sensor (ellipse method). The former method is completely passive and does not require any synchronisation. It requires at least three sensors and it is usually more accurate regarding the range estimation than azimuth/bearing estimation considering the polar coordinate system [41]. The latter method requires the synchronisation between actuation and sensing, which is available in the considered case. Also, the group velocity of propagating fastest mode must be estimated. Actuator-sensor pairs in the ellipse method become foci of an ellipse crossing damage location. However, since the distances from the foot to the sensor in the bonded ultrasonic devices are the same and expand the propagation beyond the ellipse, these distances must be subtracted (in the actual implementation, the travelling time of 37 μs multiplied by the group velocity of 5168.1 m/s estimated experimentally was subtracted). The actual foci of ellipses are actuator-foot pairs with coordinates given in Tab. 4.

The damage maps obtained by the difference in arrival method are shown in Fig. 17. Given that four sensors are available, the possible combinations without repetition of signals, namely [1, 2, 3], [1, 2, 4], [1, 3, 4], and [2, 3, 4] were utilised. For each combination, a damage map was calculated and then summed. The damage maps are presented for four simulated damage locations. It can be seen that the damage location can be estimated quite accurately, especially regarding the propagation angle measured from the centrally located actuator.

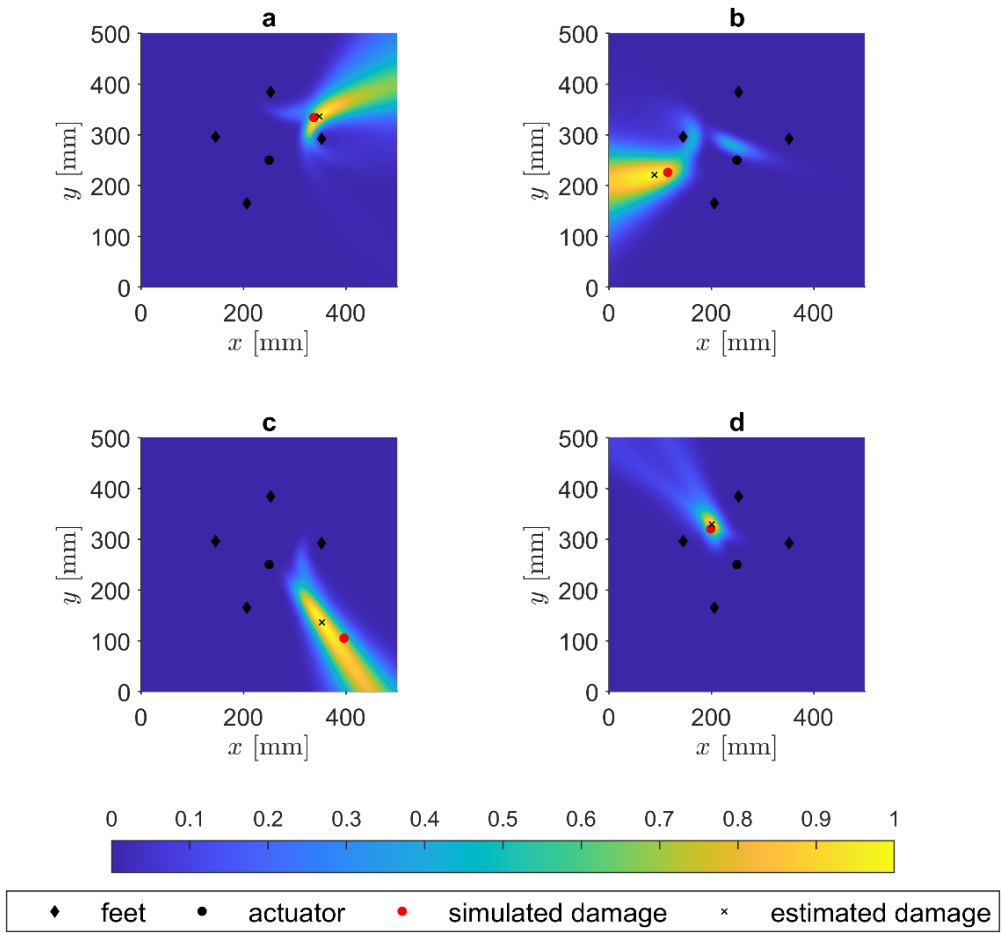

Fig. 17. Damage maps obtained by the method of difference in wave arrival (hyperbole method) for simulated damage at a) position 6, b) position 7, c) position 8, d) position 9.

Damage maps obtained by the method of time of arrival (ellipse method) are a result of the superposition of four ellipses with loci formed by actuator-feet pairs (see Fig. 18). The estimated damage locations closely align with the simulated damage. The qualitative damage assessment is superior to that achieved with the difference in arrival method. Further improvement in damage map quality can be achieved by employing more ultrasonic devices with attached piezoelectric sensors.

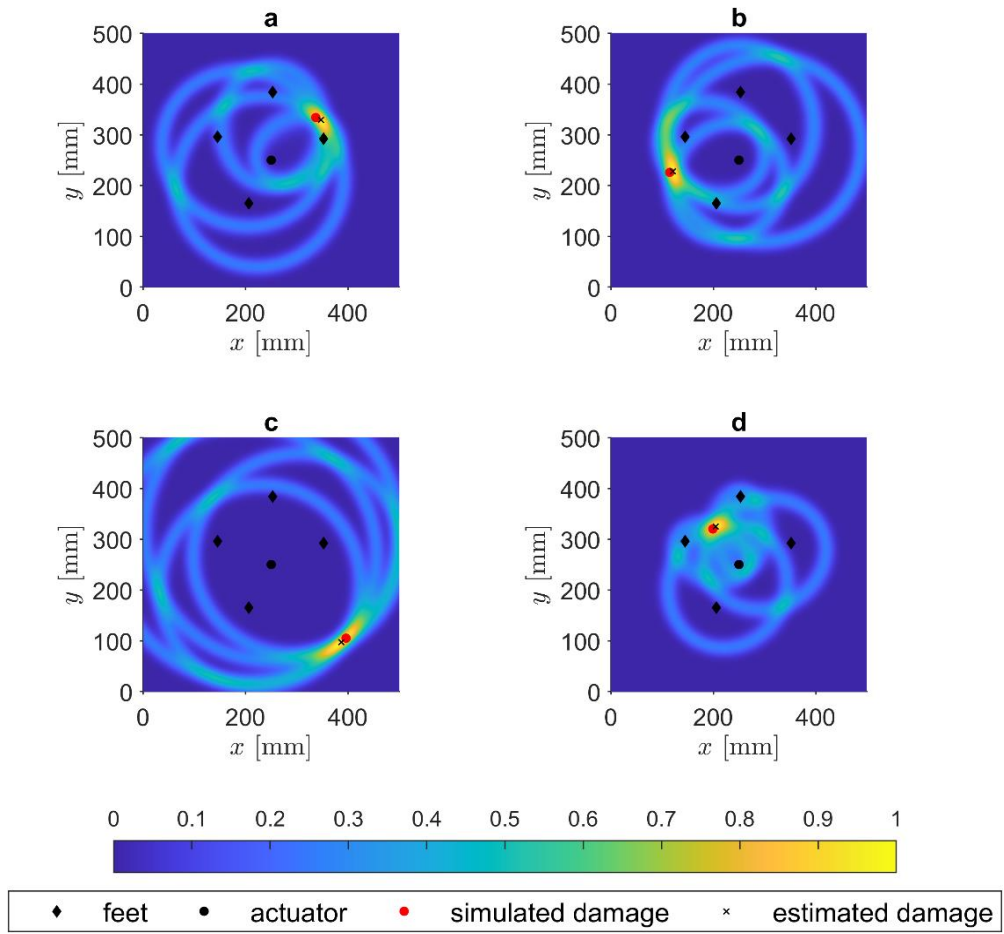

Fig. 18. Damage maps obtained by the time of arrival method (ellipse method) for simulated damage at a) position 6, b) position 7, c) position 8, d) position 9.

## 5. Conclusions

An ultrasensitive device with embedded phononic crystals was designed and tested with a focus on application in SHM. The device was inspired by spiders' legs capable of sensing vibrations and locating the prey on their web. However, instead of vibrations, nonlinear guided waves emanating from damage were utilised for damage detection and localisation. The mechanism of Bragg resonant scatterings was used for filtering the excited guided waves so that the device senses only nonlinear guided waves. Numerical and experimental tests confirmed the efficacy of the proposed method and sensitivity for higher harmonics up to the level of more than 30 times lower than the main excited harmonics.

3D printing and laser cut techniques were used for manufacturing prototypes made of polymer, aluminium, and steel. It has been found that the laser-cut steel prototype performed the best among all tested variants giving enough structural rigidity, band gap correspondence between numerical prediction and experimental measurements, and low attenuation of super-harmonics. However, experimental tests confirmed the usability of only second harmonics.

It has been shown, in an example of simulated damage, that the proposed approach can be combined with classical damage localisation algorithms based on the time of arrival and the difference in wave arrival. It has been found that the obtained damage maps allow for precise damage localisation.

In conclusion, a novel approach for damage detection and localisation has been proposed in which the device with embedded phononic crystals plays a key role. It opens new avenues in SHM. In future research, more tests are needed on real structures under dynamic loads.

**Acknowledgements**

This project has received funding from the European Union's Horizon 2020 research and innovation programme under grant agreement No 863179.

The authors would like to thank the Eko-Laser company from Bozepole Male, Poland, for providing laser-cut prototypes.